\documentclass[10pt]{article}%
\usepackage{amsfonts}
\usepackage[utf8]{inputenc}
\usepackage{natbib}
\usepackage{fullpage}
\usepackage{subfigure}
\usepackage{authblk}
\usepackage{setspace}
\usepackage{amsmath}
\usepackage{amssymb}
\usepackage{latexsym}
\usepackage{algorithm}
\usepackage{boxedminipage}
\usepackage{listings}
\usepackage{minitoc}
\usepackage{ifpdf}
\usepackage{epstopdf}
\usepackage{graphicx}%
\usepackage{RR}%

\usepackage{color}

\providecommand{\U}[1]{\protect\rule{.1in}{.1in}}
\graphicspath{{./}{./figures/}{./RR-Inria-5.1/}}

\RRNo{7895}
\RRdate{February 29, 2012}
\RRauthor{
Fran\c cois Caron\thanks{INRIA Bordeaux Sud-Ouest, Institut de Mathématiques de Bordeaux, University of Bordeaux, France}%
  \and
Luke Bornn\thanks{Department of Statistics, University of British Columbia}%
\and Arnaud Doucet\thanks{Department of Statistics, University of Oxford}}
\authorhead{F. Caron et al.}
\RRtitle{Modèles linéaires bayésiens dynamiques et parcimonieux}
\RRetitle{Sparsity-Promoting Bayesian Dynamic Linear Models}
\titlehead{Sparsity-Promoting Bayesian Dynamic Linear Models}
\RRresume{Les distributions a priori encourageant la parcimonie sont devenues de plus en plus populaires au cours des dernières années du fait du nombre d'applications croissantes en régression et classification impliquant un grand nombre de prédicteurs. Dans le cas où les observations sont recueillies au cours du temps, il est souvent irréaliste de considérer que la structure de parcimonie est fixée au cours du temps. Nous proposons ici une classe originale de modèles bayésiens linéaires flexibles pour la modélisation dynamique parcimonieuse. La classe de modèles proposée repose sur l'utilisation de distributions hyperboliques généralisées. Les propriétés de ces modèles sont explorées au travers de résultats analytiques et de simulations. Enfin, nous présentons une application de ce modèle en finance.}
\RRabstract{
Sparsity-promoting priors have become increasingly popular over recent years
due to an increased number of regression and classification applications
involving a large number of predictors. In time series applications where
observations are collected over time, it is often unrealistic to assume that
the underlying sparsity pattern is fixed. We propose here an original class of
flexible Bayesian linear models for dynamic sparsity modelling. The proposed
class of models expands upon the existing Bayesian literature on sparse
regression using generalized multivariate hyperbolic distributions. The
properties of the models are explored through both analytic results and
simulation studies. We demonstrate the model on a financial application where
it is shown that it accurately represents the patterns seen in the analysis of
stock and derivative data, and is able to detect major events by filtering an
artificial portfolio of assets. }
\RRmotcle{Modèles parcimonieux, distribution hyperbolique généralisée, modèle de mélange de gaussiennes, régression linéaire dynamique}
\RRkeyword{generalized hyperbolic, Gaussian mixture models, sparsity, dynamic regression}
\RRprojet{ALEA}

\RCBordeaux 

\begin{document}

\makeRR   

\section{Introduction}

Over recent years, there has been an increased number of regression and
classification applications involving high-dimensional data. In these
scenarios, it is common to have a large number of predictors, a number of them
being irrelevant. The need to appropriately restricts the number of predictors
for improved statistical efficiency and predictive abilities has generated a
large body of work. In the non-Bayesian literature sparse regression analysis
via penalised likelihood has become extremely popular since the seminal
lasso\ paper~\citep*{Tibshirani1996}; in the Bayesian literature, spike-and-slab priors
have historically been favoured \citep*{Mitchell1988,Zhang2007}.
Unfortunately, spike and slab priors are notoriously difficult to fit, leading to a renewed interest in proposing alternative sparsity-promoting prior models.

It is well-known that the Lasso estimate for linear regression parameters can
be interpreted as the MAP\ (Maximum A\ Posteriori) estimate when the
regression parameters are assigned independent Laplace priors. From a Bayesian
perspective, the use of MAP estimators lacks solid justification; however, although they are not sparse in the exact sense, Bayesian posterior medians are remarkably similar in value to lasso
estimates \citep*{Park2008} and provide credible intervals which can help in guiding variable selection. Additionally it has been observed empirically
that Markov chain Monte Carlo (MCMC) mix quite well for such models
\citep*{Kyung2010}. However, it is well-known that the lasso estimates and its Bayesian
version suffer from various problems. In particular, coefficients can get
shrunk towards zero even when there is overwhelming evidence in the likelihood
that they are non-zero. There has been much work in the non-Bayesian and
Bayesian literature to improve over this; for example, a number of
sparsity-promoting non-concave log prior distributions have been proposed
which reduce bias in the estimates of large coefficients. Recent work includes
\citep*{Griffin2007, Caron2008, Lee2010,Griffin2010} and
Bayesian interpretations of the group lasso and elastic net estimators
\citep*{Bornn2010,Li2010,Kyung2010}. Although the above priors result in
non-sparse posterior median and mean estimates, many arguments have been
advocated in favour of their use, see e.g. \citep*{Kyung2010}.

In the context of time series, it is of particular interest to allow for the
sparsity pattern to evolve over time as a predictor which is highly relevant
in a given time period may become irrelevant later on. Dynamic sparsity
modelling is an important topic that has received much less attention in the
literature. From a non-Bayesian perspective, several authors have proposed to
adapt the elastic net, fused lasso or group lasso to accommodate dynamic models
\citep*{Angelosante2009,Angelosante2009a,Vaswani2008,Jacob2009}. From a Bayesian
perspective, dynamic spike-and-slab type models have been recently proposed:
\cite{Nakajima2011} associate to each predictor a latent process and this predictor
is only included in the regression when the magnitude of its associated latent
process is above a given latent threshold, whereas \cite{Ziniel2010} associate
to each predictor a latent binary inclusion/exclusion Markov chain. We follow
here an alternative approach based on the construction of dynamic
sparsity-promoting priors. Similar constructions were also recently proposed
independently in \citep{Sejdinovic2010} and \citep{Kalli2012}. However, the model presented here is
much more flexible. It relies on a scale mixture of normal distributions
where the mixing distribution is itself a generalized inverse Gaussian
resulting in a multivariate generalized hyperbolic distribution. This scale
mixture of normals representation can be used to derive tailored Markov chain
Monte Carlo (MCMC) and Sequential Monte Carlo (SMC) methods for inference.

The rest of the paper is organized as follows. In Section~\ref{sec:sparsebayes}, we review the
generalized hyperbolic distribution and show that it includes numerous
sparsity-promoting priors used in the literature. Section~\ref{sec:dynamicsparsebayes} presents our
dynamic sparsity model and establishes some of its properties. Section~\ref{sec:algorithms}
proposes several Bayesian computational procedures to perform inference. We demonstrate the model on simulated data and an application to
financial data in Sections~\ref{sec:simulation} and~\ref{sec:stockvolatility}.

\section{Sparse Bayesian regression}
\label{sec:sparsebayes}
\subsection{Bayesian regression model}

Consider the following standard regression model where
\begin{equation}
y=X\beta+\epsilon
\end{equation}
where $y$ is the $n\times1$ vector of responses, $\beta=\left(  \beta
_{1},...,\beta_{p}\right)  ^{\text{T}}$ the vector of regression parameters,
$X$ is the $n\times p$ design matrix, and $\epsilon$ is the $n\times1$ vector
of independent and identically distributed normal errors with mean $0$ and
variance $\sigma^{2}$. In a Bayesian approach, we adopt a prior density
$\pi\left(  \beta\right)=\prod_{j=1}^p \pi\left(  \beta_j\right)$. \cite{Park2008} proposed to use independent Laplace priors, motivated by the fact that lasso estimates could be interpreted as the Bayes posterior mode under this prior~\citep*{Tibshirani1996}. Laplace priors can be expressed as scale mixture of normal distributions~\citep*{Andrews1974,West1987}, hence admit a hierarchical construction. Other models, based on scale mixture of normals, have been proposed in the literature~\citep*{Tipping2001,Caron2008,Griffin2010}. Several of these distributions can be considered as particular case of the generalized hyperbolic distribution, which we review in the next section.

\subsection{Generalized hyperbolic distribution}

Let $\beta_{j}\in\mathbb{R}$, and suppose the following Gaussian mixture
model
\begin{align}
\beta_{j}|\tau_{j}  &  \sim\mathcal{N}(\mu,\tau_{j})\\
\tau_{j}  &  \sim GiGauss(\nu,\delta,\gamma)
\end{align}
where $\mathcal{N}(\mu,\sigma^{2})$ denotes the Gaussian distribution of mean
$\mu$ and variance $\sigma^{2}$ and $GiGauss(\nu,\delta,\gamma)$ is the
generalized inverse Gaussian distribution \citep*{Barndorff-Nielsen2001} of
parameters $\nu,\delta,\gamma$ whose probability density function is%
\begin{equation}
\frac{(\gamma/\delta)^{\nu}}{2K_{\nu}(\delta\gamma)}x^{\nu-1}\exp\left(
-\frac{1}{2}(\delta^{2}x^{-1}+\gamma^{2}x)\right)  \text{, }x>0
\end{equation}

$\beta_{j}$ then follows a generalized hyperbolic distribution of pdf
\begin{equation}
\frac{(\gamma/\delta)^{\nu}}{\sqrt{2\pi}\gamma^{\nu-1/2}K_{\nu}(\delta\gamma
)}\left (\sqrt{\delta^{2}+(\beta_{j}-\mu)^{2}}\right )^{\nu-1/2}K_{\nu-1/2}\left (\gamma \sqrt{\delta^{2}+(\beta_{j}-\mu)^{2}}\right )
\end{equation}
where $K_{a}(z)$ is the modified
Bessel function of the third kind. We write $\beta_{j}\sim GH(\mu,\nu
,\delta,\gamma)$. When $\mu=0$, the distribution is concentrated around $0$.
For some values of the parameters $\nu$, $\delta$ and $\gamma$, the pdf will
be concentrated around 0 with heavy tails, which makes it a desirable prior
distribution for sparse linear regression. The generalized hyperbolic
distribution generalizes several distributions that have been used as
sparsity-promoting priors:

\begin{enumerate}
\item[(a)] $\nu=-1/2$: Normal inverse Gaussian law \citep*{Caron2008}

\item[(b)] $\delta=0$, $\nu>0$: Normal gamma law
\citep*{Caron2008,Griffin2010}

\item[(c)] $\delta=0$, $\nu=1$: Laplace law \citep*{Tibshirani1996,Park2008}

\item[(d)] $\gamma=0$, $\nu<0$: Student's law~\citep*{Tipping2001}
\end{enumerate}

All of these priors have been described as suitable prior distributions for
promoting sparsity/shrinkage in Bayesian regression models. In some cases,
they have been shown to give better predictive performances than
spike-and-slab priors in regression~\citep{Griffin2010}. The mixing properties of the associated Markov chain Monte Carlo algorithms are also generally
considered to be superior due to the smooth scale mixture representation.

\begin{figure}[ptb]
\begin{center}
\subfigure[]{\includegraphics[width=6cm]{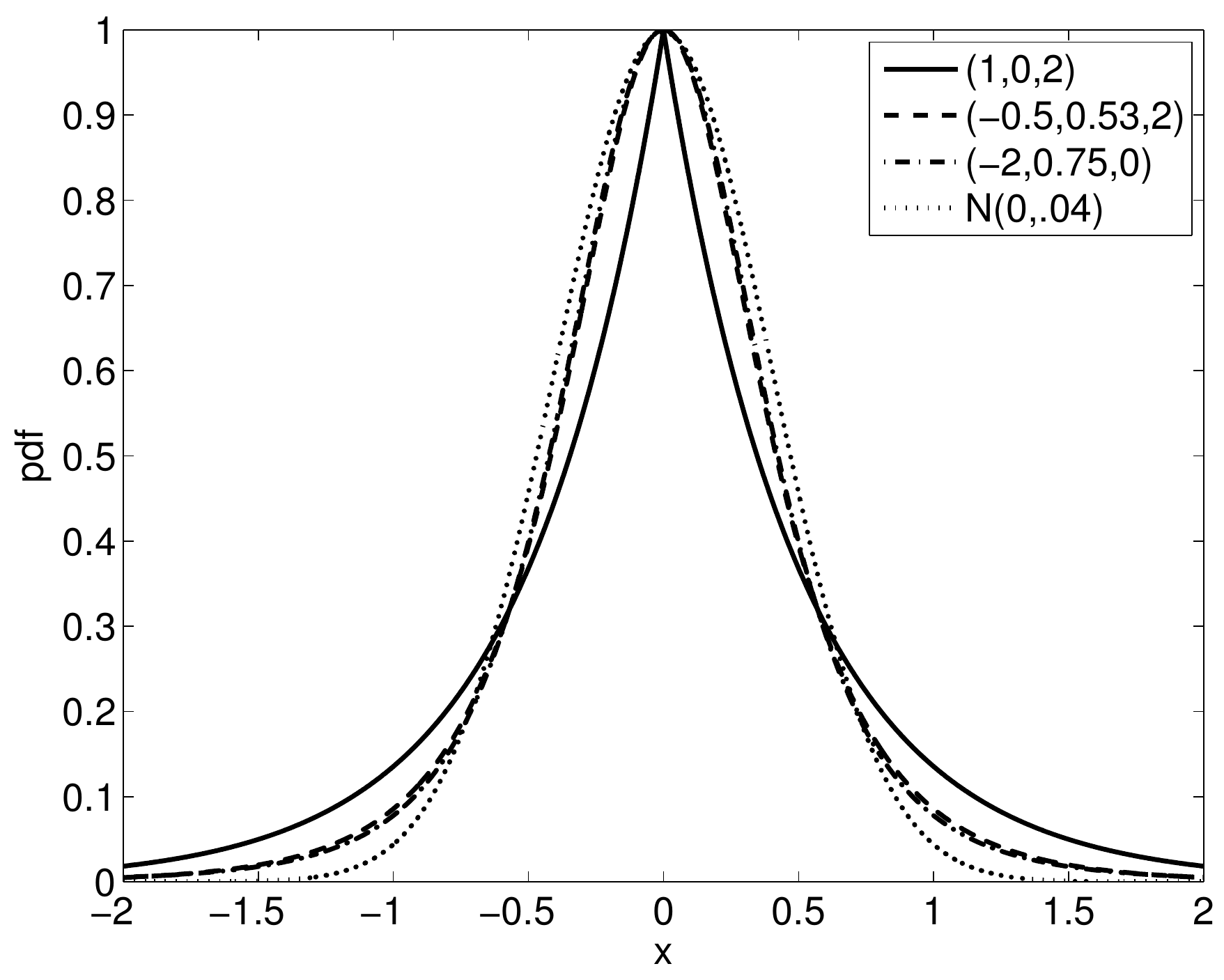}}
\subfigure[]{\includegraphics[width=6cm]{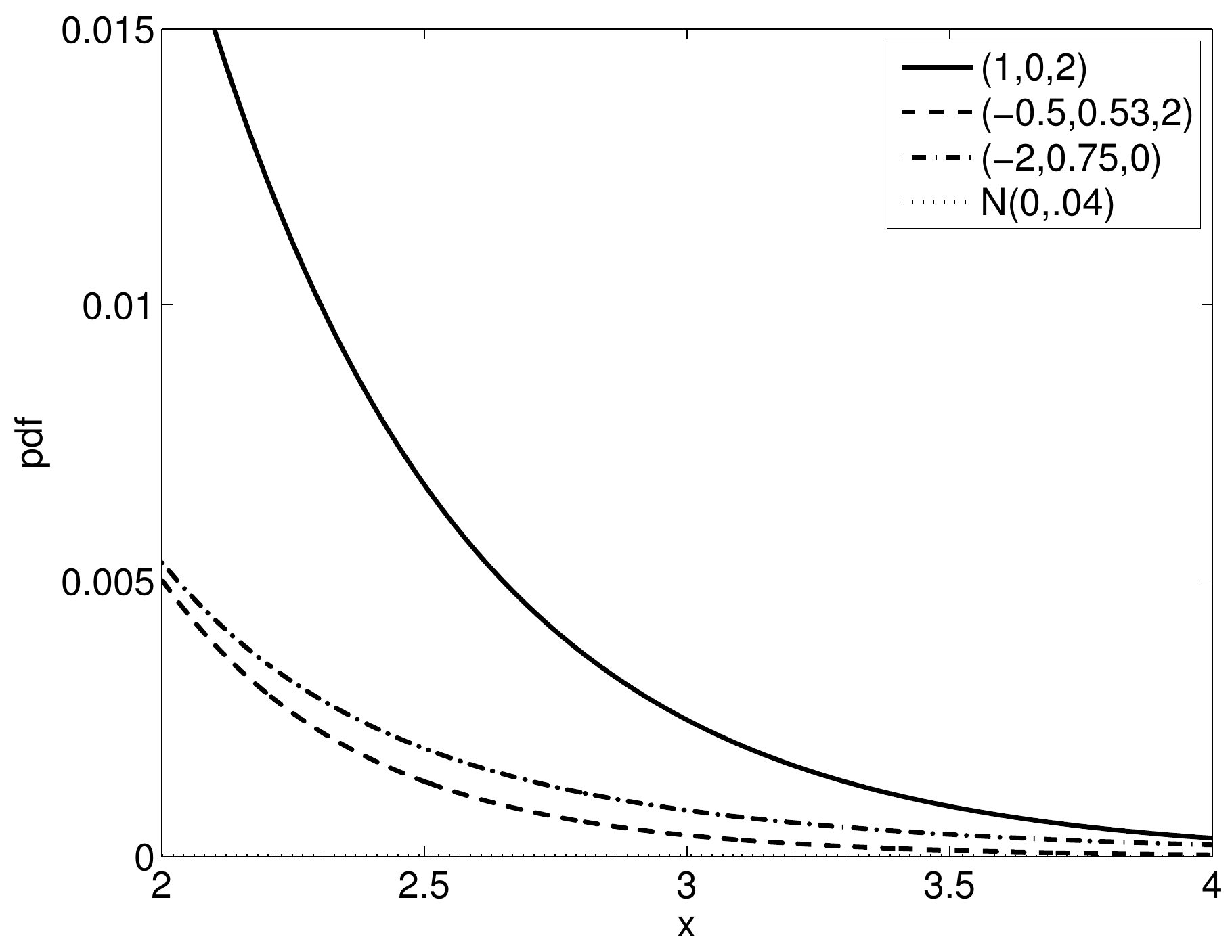}}
\end{center}
\caption{Probability density functions of the generalized hyperbolic
distribution for several values of $(\nu,\delta,\gamma)$ correspond to a
Laplace, Normal inverse Gaussian and Student t distribution. The pdf of the
normal distribution is also shown for comparison. (a) Behavior around 0 and
(b) Tail behavior.}%
\label{fig:pdf}%
\end{figure}

The probability density
functions of the generalized hyperbolic distribution for different values of
the parameters are represented in Figure~\ref{fig:pdf}, where we see that the
parameters $\nu$, $\delta$ and $\gamma$ provide significant control over the
behavior of the mode and tails.

\section{Dynamic Sparse Bayesian regression}
\label{sec:dynamicsparsebayes}

We now consider the problem of successive linear regression models%
\begin{equation}
y_{t}=X_{t}\beta_{t}+\varepsilon_{t}\label{eq:linreg}%
\end{equation}
where $t=1,\ldots,T$ is a time index, $\varepsilon_{t}\sim\mathcal{N}%
(0,\sigma^{2}I_{n})$, $X_{t}$ is a $n\times p$ design matrix, $I_{n}$ is
the $n\times n$ identity matrix and $\beta_{t}\in\mathbb{R}^{p}$.  Assume a priori
independence of the different components $j=1,\ldots,p$%
\begin{equation}
\pi(\beta_{1:T})=\prod_{j=1}^{p}\pi(\beta_{j,1:T})
\end{equation}
We assume
that the true vector $\beta_{t}$ is sparse, and that the sparsity pattern
(indices of elements with zero values) is slowly evolving over time. We first consider a simple particular case of group sparsity where the sparsity pattern is unchanged over time, and introduce the multivariate hypergeometric distribution. We then describe how to use this distribution to define models where the sparsity pattern evolves smoothly over time, while highlighting several interesting statistical properties.

\subsection{Multivariate generalized hyperbolic}

Assume as a simple starting case that we want the sparsity pattern to be shared across time. This can be achieved by considering a hierarchical Gaussian model where the variance is shared over time. Suppose that, for $t=1,\ldots,T$
\begin{equation}
\beta_{j,t}|\tau_j\sim\mathcal{N}(\mu,\tau_j\Sigma)
\end{equation}
where $\mu\in\mathbb{R}^{p}$, $\Sigma$ is a positive semi-definite $p\times p$
matrix, and $\tau_j\sim GiGauss(\nu,\delta,\gamma)$. Then $\beta_{j,1:T}\in
\mathbb{R}^{T}$ follows the multivariate generalized hyperbolic distribution
of pdf%
\begin{equation}
\frac{(\gamma/\delta)^{\nu}}{(2\pi)^{p/2}\gamma^{\nu-p/2}K_{\nu}(\delta
\gamma)}q^{\nu-p/2}K_{\nu-p/2}(\gamma q)
\end{equation}
where $q=\sqrt{\delta^{2}+(\beta_{j,1:T}-\mu)^{T}\Sigma^{-1}(\beta_{j,1:T}-\mu)}$. We write
$\beta_{j,1:T}\sim mGH(\mu,\nu,\delta,\gamma,\Sigma)$. Shared sparsity pattern over
the $\beta_{j,t}^{\prime}s$ is obtained through the shared variance term $\tau_j$.
The matrix $\Sigma$ allows one to introduce correlation between variables;
again, such priors have been studied in the literature on sparse models. In
particular, the group lasso prior \citep*{Yuan2006,Raman2009,Kyung2010} defined by
\begin{equation}
\pi(\beta_{j,1:T})\propto\exp(-\gamma\left\Vert \beta\right\Vert _{\Sigma
})\label{eq:GroupLasso}
\end{equation}
is a special case of the multivariate hyperbolic when $\nu=\frac{p+1}{2}$,
$\mu$ is the null vector and $\delta=0.$ Other special cases such as the
multivariate normal gamma and normal inverse Gaussian have also been studied
by \citet*{Caron2008}.

\subsection{Statistical model}

We now turn to the use of the multivariate generalized hyperbolic distribution
in the modeling of data with dependent and varying sparsity structure. We are interested in defining a model for $\pi(\beta_{j,1:T})$ that introduces
correlations in time both
\begin{enumerate}
\item[(a)] in the sparsity pattern: if the vector of regressors is sparse at time $t$, then
it is more likely to be sparse at time $t+1$, and

\item[(b)] in the value of non-zero coefficients.
\end{enumerate}

These properties will be obtained by considering a particular decomposition of the joint distribution $\pi(\beta_{j,1:T})$ with multiple overlapping groups. Consider the following decomposition of the joint distribution:
\begin{equation}
\pi(\beta_{j,1:T})=\frac{\prod_{t=d}^{T}\pi(\beta_{j,t-d:t})}{\prod
_{t=d}^{T-1}\pi(\beta_{j,t-d+1:t})}%
\end{equation}
where $\beta_{j,t-d:t}$ is marginally $mGH(0_{d+1},\nu,\delta,\gamma
,\Sigma_{d+1})$, where $d>0$, $0_{d}$ is the null vector of length $d$, $$\Sigma_{d}=\left(
\begin{array}
[c]{ccccc}%
1 & \alpha &  \ldots & \alpha^{d-1}\\
\alpha & 1 &  \ldots & \ldots\\
\ldots & \ldots & \ldots & \ldots \\
\alpha^{d-1} & \alpha^{d-1} & \ldots &  1
\end{array}
\right)  $$ and $\alpha\in\lbrack0,1]$. In particular, if $\nu=\frac{d+2}{2}$ and $\delta=0$, then
$\beta_{j,t-d:t}$ follows the group lasso distribution~\eqref{eq:GroupLasso}.
\bigskip

The model can be alternatively defined by the following d-order Markov model ($d>0$)%
\begin{equation}
\beta_{j,1:d}\sim mGH(0_{d},\nu,\delta,\gamma,\Sigma_{d})\label{eq:init}%
\end{equation}
and for $t>d$%
\begin{equation}
\beta_{j,t}|\beta_{j,1:t-1}\sim GH(\alpha\beta_{j,t-1},\nu-d/2,\sqrt
{1-\alpha^{2}}\sqrt{\delta^{2}+\left\Vert \beta_{j,t-d:t-1}\right\Vert
_{\Sigma_{d}}^{2}},\frac{\gamma}{\sqrt{1-\alpha^{2}}})\label{eq:predictive1}%
\end{equation}
where $\left\Vert x\right\Vert _{\Sigma
}=\sqrt{x^{T}\Sigma^{-1}x}$ is the Mahalanobis distance.\ In the case $d=0$,
$\beta_{j,t}$ are iid $GH(0,\nu,\delta,\gamma)$. Using the scale mixture representation of the generalized hyperbolic distribution, the predictive distribution
(\ref{eq:predictive1}) can be equivalently expressed as a scale mixture of
normals with latent variables $\tau_{j,t}$
\begin{align*}
\tau_{j,t}|\beta_{j,1:t-1}  &  \sim GiGauss\left(  \nu-d/2,\sqrt{\delta
^{2}+\left\Vert \beta_{j,t-d:t-1}\right\Vert _{\Sigma_{d}}^{2}},\gamma\right)
\\
\beta_{j,t}|\tau_{j,t},\beta_{j,t-1}  &  \sim\mathcal{N}\left(  \alpha
\beta_{j,t-1},(1-\alpha^{2})\tau_{j,t}\right).
\end{align*}

The model has the following statistical properties:

\begin{enumerate}
\item[(a)] The model is first-order stationary with%
\[
\beta_{j,t}\sim GH(0,\nu,\delta,\gamma)
\]

\item[(b)] For any $h\leq d$%
\[
\beta_{j,t-h:t}\sim mGH(0_{h+1},\nu,\delta,\gamma,\Sigma_{h+1})
\]

\item[(c)] The parameter $0\leq d\leq T-1$ tunes the evolution of the sparsity
pattern over time, and we have the following special cases

\begin{itemize}
\item $d=0$, we have independence between the sparsity patterns over time and
$\beta_{j,t}$ are iid $GH(0,\nu,\delta,\gamma)$

\item $d=T-1$, the sparsity pattern is shared over time and $\beta_{j,1:T}\sim
mGH(0_{T},\nu,\delta,\gamma,\Sigma_{T})$
\end{itemize}

\item[(d)] The parameter $\alpha\in\lbrack0,1]$ tunes the correlation between
regression coefficient values at successive time steps.\bigskip
\end{enumerate}

\begin{figure}[ptb]
\begin{center}
\subfigure[$d=5$]{\includegraphics[width=7.5cm]{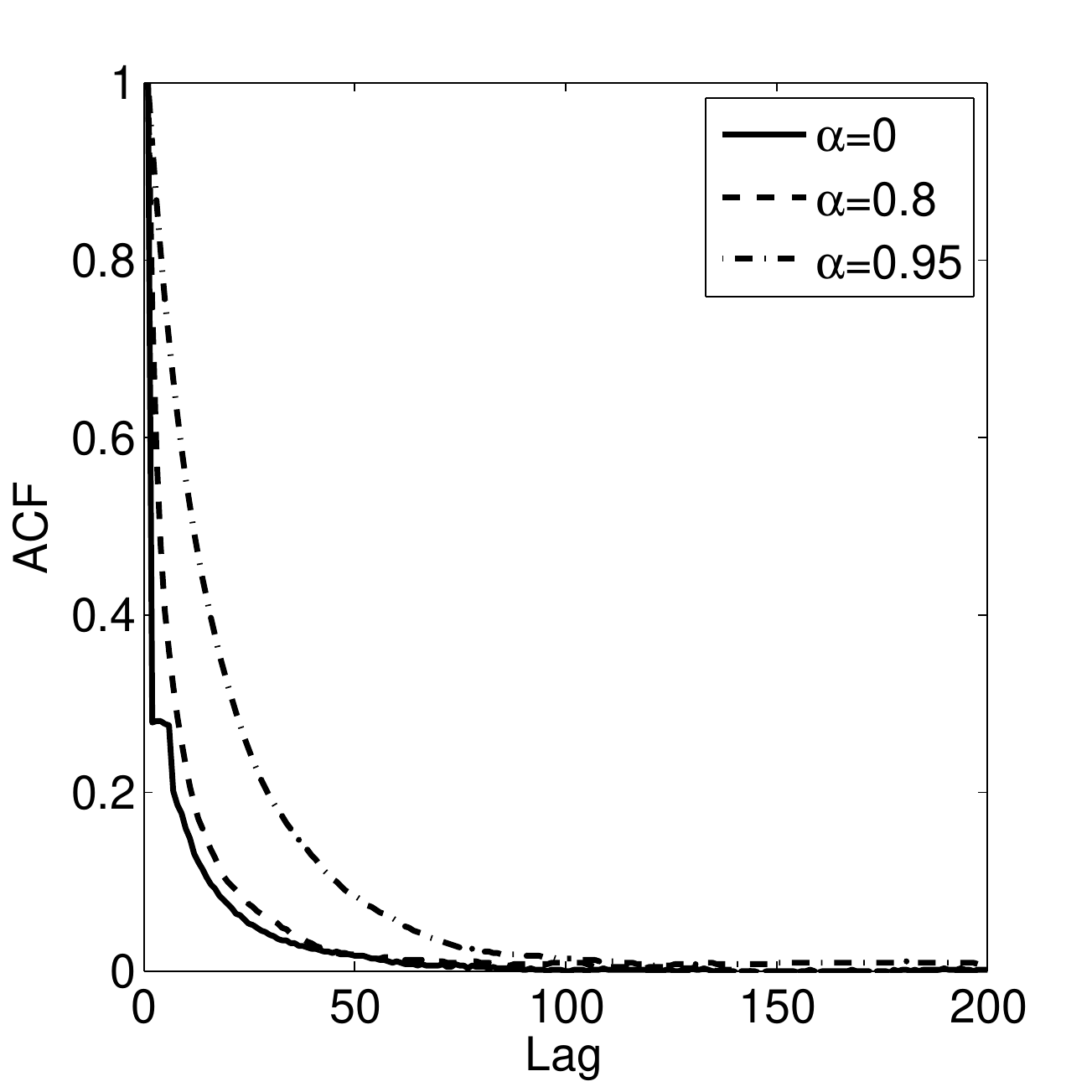}}
\subfigure[$d=20$]{\includegraphics[width=7.5cm]{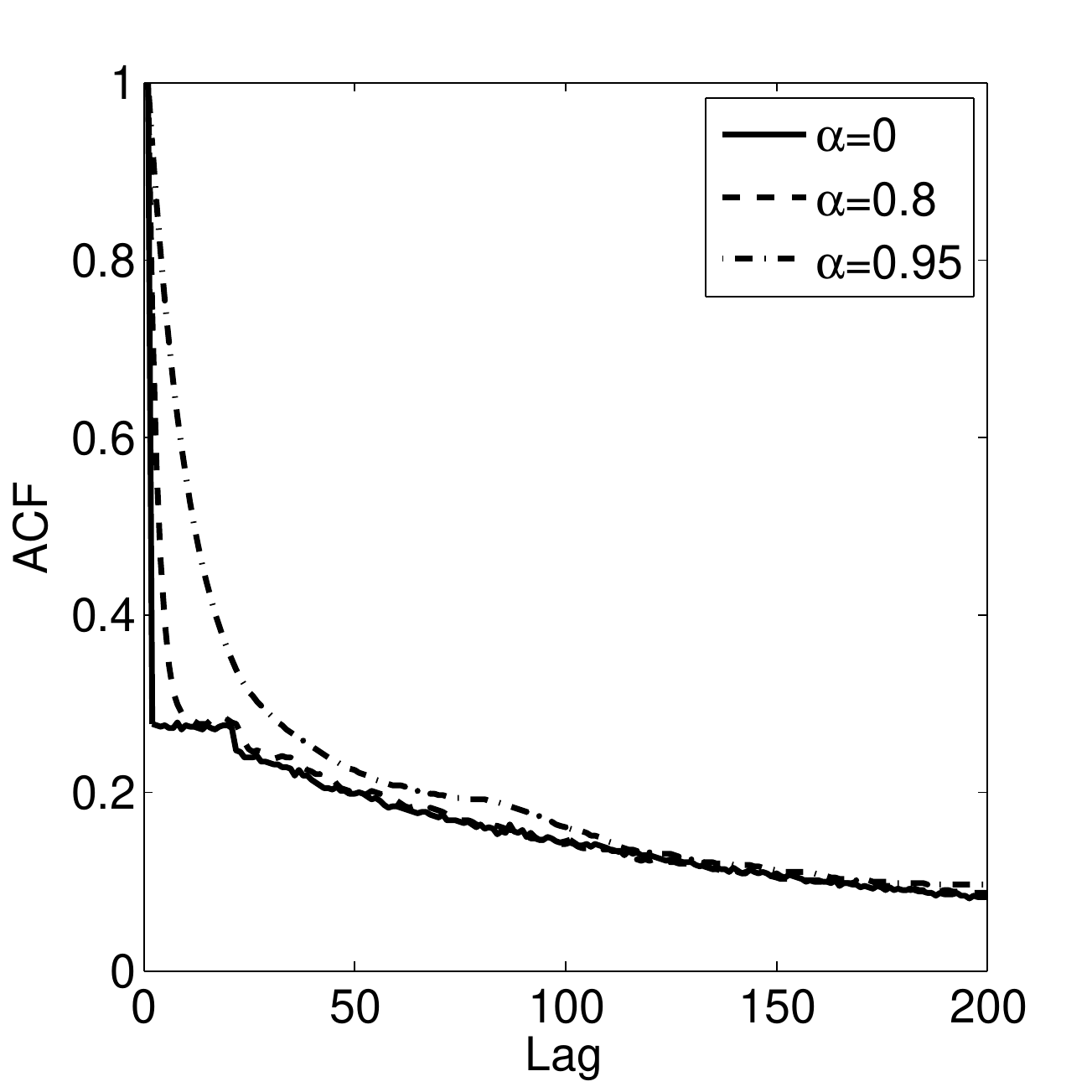}}
\end{center}
\caption{Autocorrelation function for $\beta_{t}^{2}$ from the statistical
model defined by Eq. \eqref{eq:init} and \eqref{eq:predictive1} with $\nu
=0.1$, $\delta=0.01$, $\gamma=1$, $T=10^{6}$ and (a) $d=5$, (b) $d=20$. For
each value of $d$, three plots are represented with $\alpha=0.00, 0.80, 0.95$.
Note that $d=0$ is not represented, since the variables $\beta_{t}$ are
independent in that case. (b) For $d=20$, we can clearly see that the
correlation for $\beta_{t}^{2}$ is due to both $\alpha$ and the shared
sparsity pattern; after lag 100, the correlation remains due to the shared
sparsity pattern. For $\alpha=0$ (black line) we can clearly see a threshold on autocorrelation, which is due to the shared sparsity pattern induced by the model.}%
\label{fig:corr1}%
\end{figure}

\begin{figure}[ptb]
\begin{center}
\includegraphics[width=7.5cm]{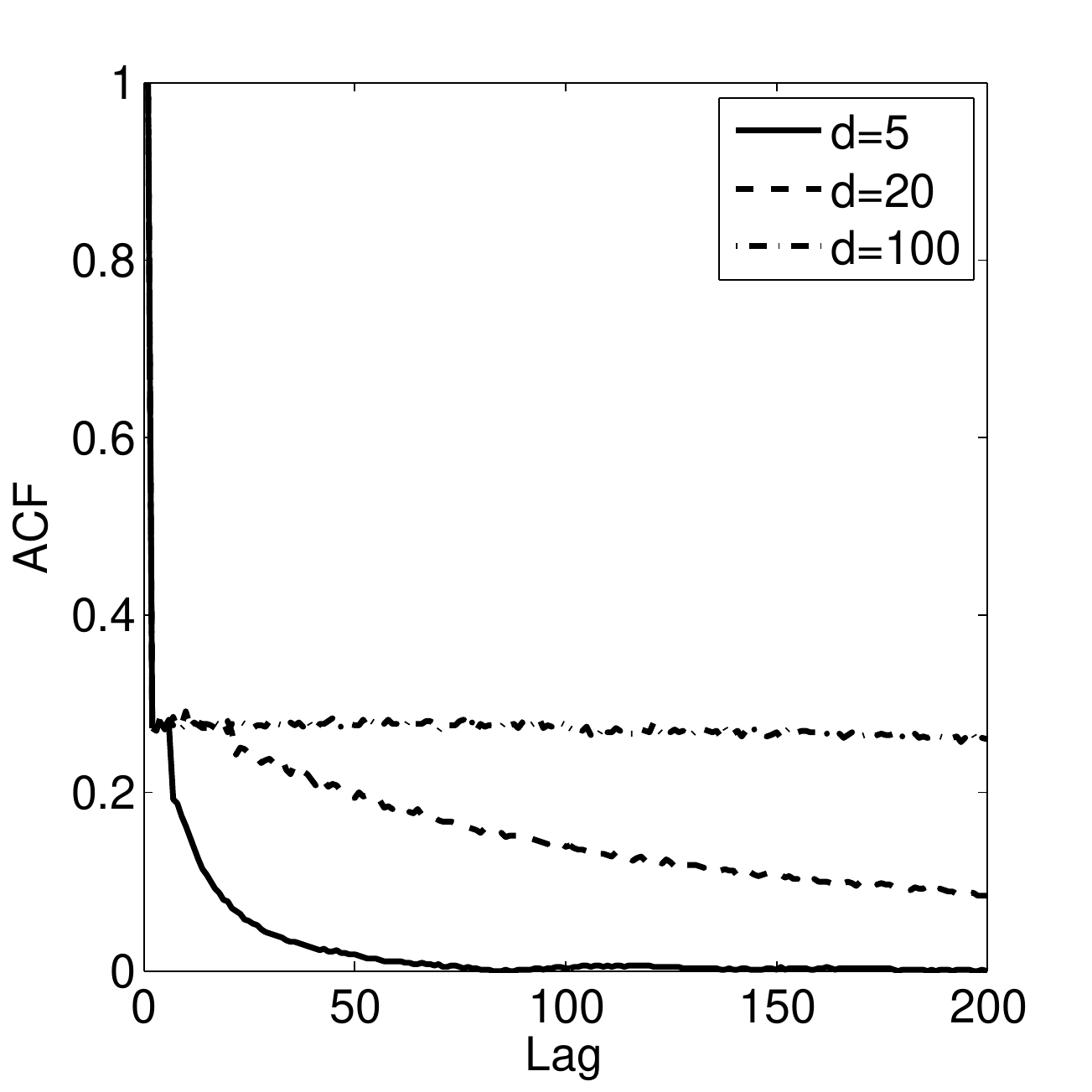}
\end{center}
\caption{Autocorrelation function for $\beta_{t}^{2}$ from the statistical
model defined by Eq. \eqref{eq:init} and \eqref{eq:predictive1} with $\nu
=0.1$, $\delta=0.01$, $\gamma=1$, $T=10^{6}$ and $\alpha=0$. Three plots are
represented with $d=5, 20, 100$. As $\alpha=0$, the samples $(\beta_{t})$ are
uncorrelated, and the autocorrelation for $\beta_{t}^{2}$ is due to the shared
sparsity pattern.}%
\label{fig:corr2}%
\end{figure}

These properties make the model very appealing for dynamic linear regression.
First, by choosing the parameters $\nu$, $\delta$ and $\gamma$ based on the
large literature on sparse Bayesian regression, the user can define the level
of sparsity desired in the signal. For example, if $\delta=0 $ and $\nu>0$
(normal-gamma case) smaller values of $\nu$ will favor sparser solutions.
Second, the user will define how this sparsity pattern is going to evolve over
time, from the two extreme cases $d=0$ (independent sparsity pattern over
time) and $d=T$ (shared sparsity). Between those two extremes, the value of
$d$ will tune how often the time series can alternate between sparse and
non-sparse periods. This effect can be seen by looking at the autocorrelation
plot for $\beta_{t}^{2}$, as shown in figures~\ref{fig:corr1}
and~\ref{fig:corr2}. The shared sparsity pattern induces a minimum level of autocorrelation over the lag $d$, as can be seen from Figure \ref{fig:corr2} for $\alpha=0$. Finally, the parameter $\alpha$ tunes the correlation
between non-zero coefficients, in a classical way, and we have $\text{corr}%
(\beta_{t},\beta_{t-1})=\alpha$. 

\begin{figure}[ptb]
\begin{center}
\subfigure[$d=0$]{\includegraphics[width=8cm]{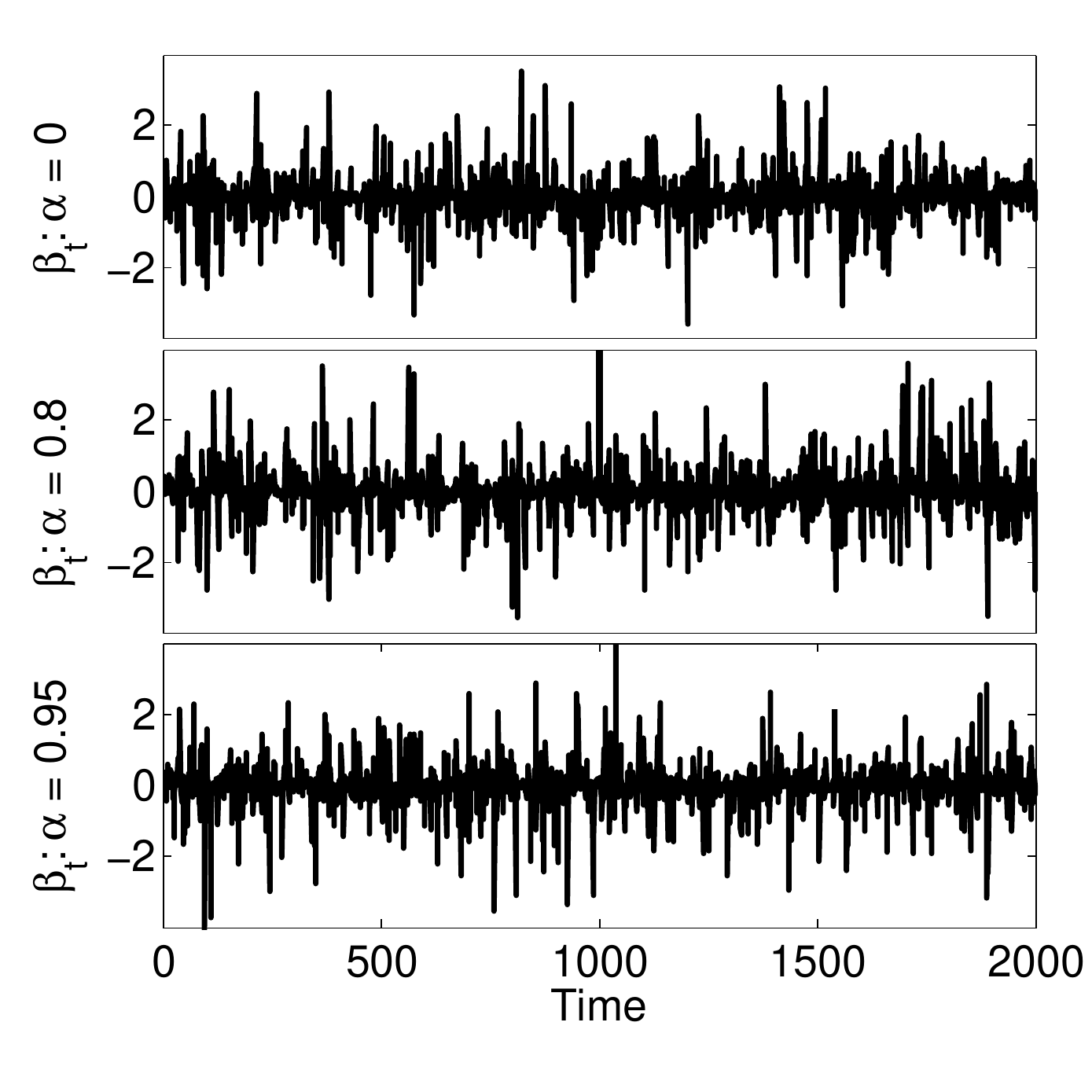}}
\subfigure[$d=5$]{\includegraphics[width=8cm]{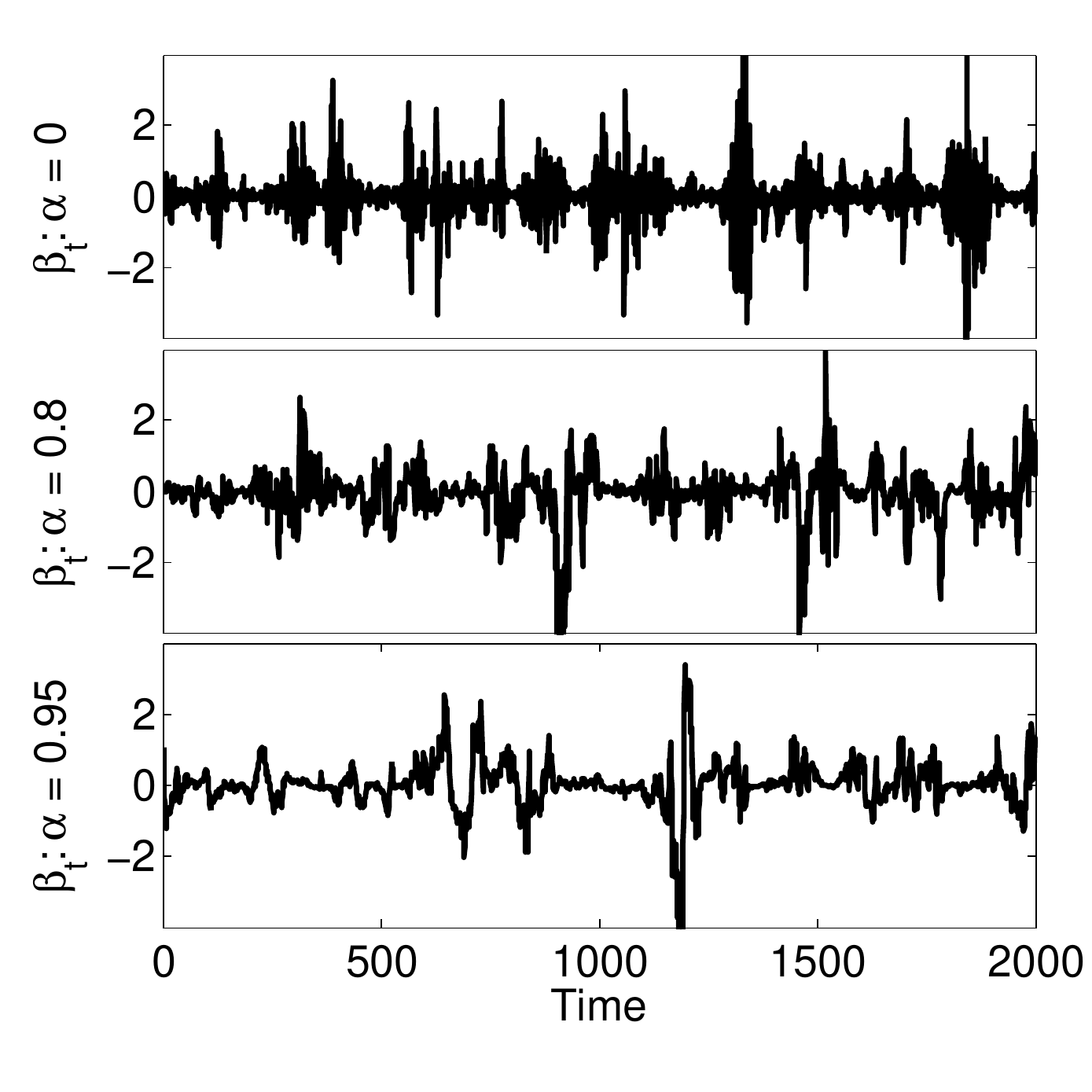}}
\subfigure[$d=20$]{\includegraphics[width=8cm]{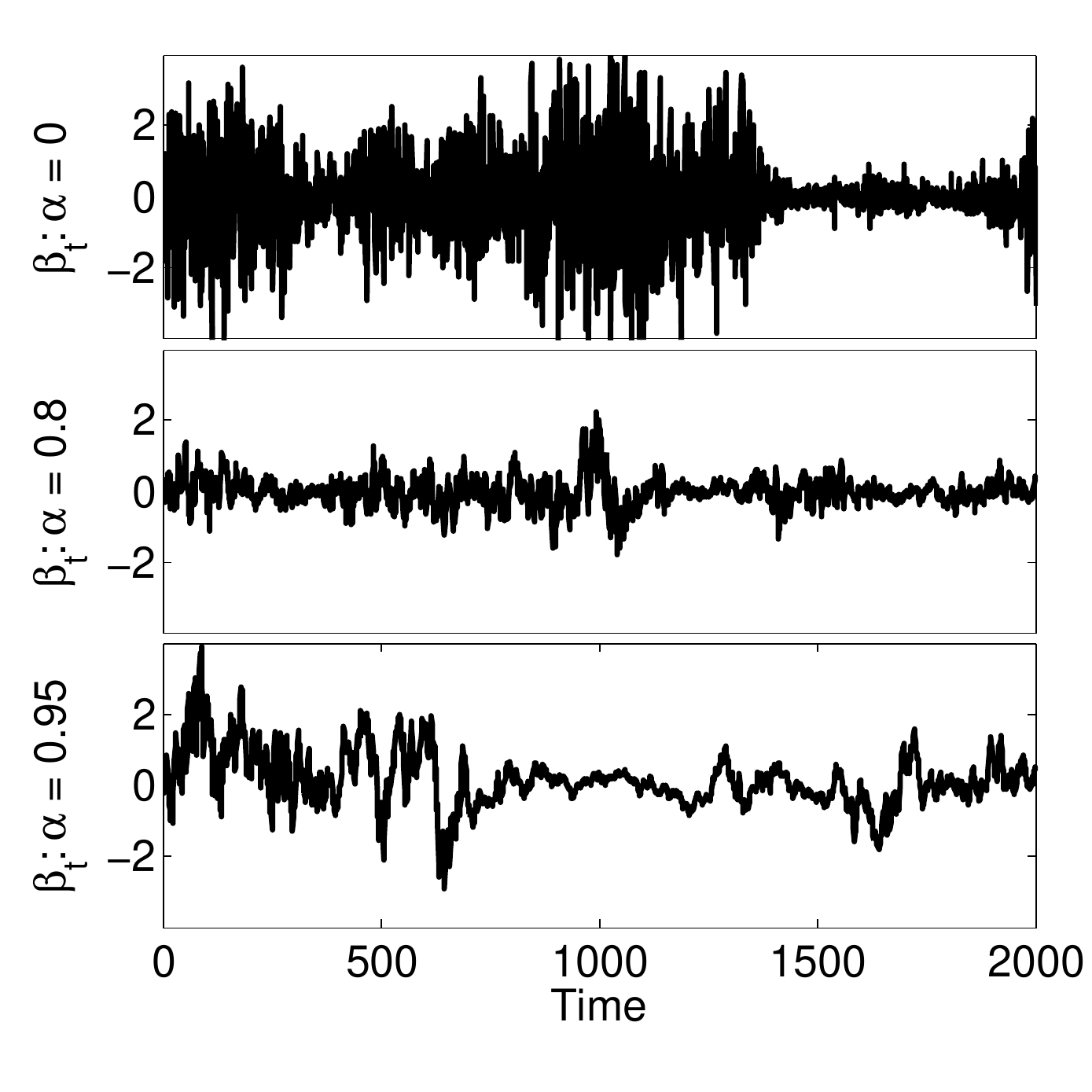}}
\subfigure[$d=2000$]{\includegraphics[width=8cm]{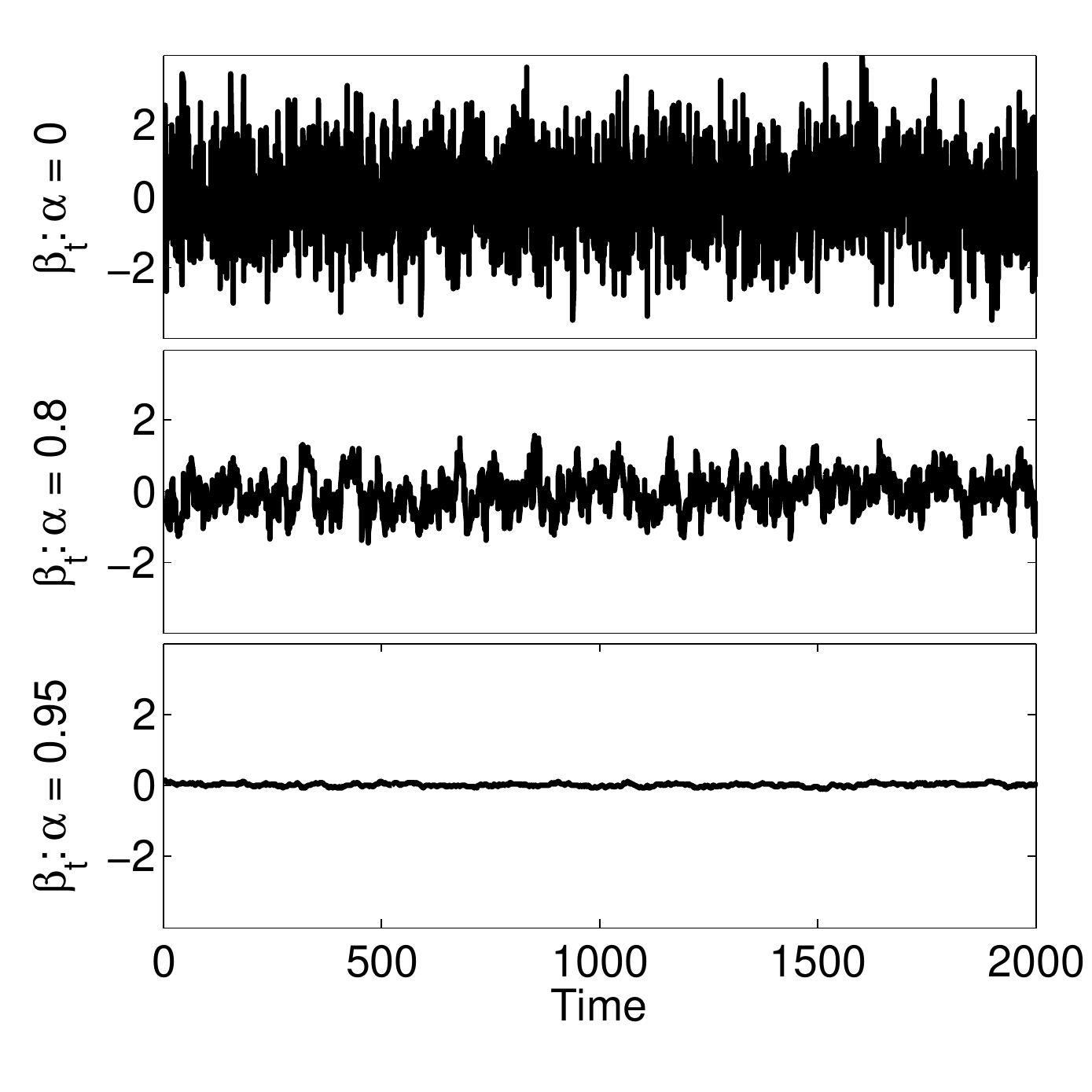}}
\end{center}
\caption{Samples from the statistical model defined by Eq. \eqref{eq:init} and
\eqref{eq:predictive1} with $\nu=0.1$, $\delta=0.01$, $\gamma=1$, $T=2000$ and
(a) $d=0$, (b) $d=5$, (c) $d=20$, (d) $d=T$. For each value of $d$, three
draws are represented with $\alpha=0.00, 0.80, 0.95$. (a) For $d=0$, the
variables $\beta_{t}$ are independent and the sparsity pattern is not shared.
(b,c) For $d=5, 20$, the sparsity pattern can evolve over time, and the
process alternates regions close to zero and away from zero. (d) For $d=T$,
the sparsity pattern is shared, and the $\beta_{t}$ are either all close to
zero (bottom figure in (d)) or away from zero (top and middle figure in (d)).}%
\label{fig:samples}%
\end{figure}

In Figure~\ref{fig:samples}, we represent some samples from this model for
different values of $d$ and $\alpha$, to show how the sparsity pattern evolves
over time depending on this parameter. This figure motivates the model for
use in the modeling of stock volatility. Specifically, stock prices (and
trading activity) go through alternating periods of inactivity and alacrity,
making the multivariate generalized hyperbolic distribution a suitable
modeling choice.

The model defined by Equations (\ref{eq:init}) and (\ref{eq:predictive1})
relies on an integer $d$ that tunes the evolution of the sparsity pattern. One
might want to consider this parameter as varying over time, and estimate this
value. We could then put a prior on $d_{t}$, e.g. a Markov model%
\[
d_{t}|d_{t-1}\sim Bin(d_{t-1}+1,\rho)
\]
where $\rho\in\lbrack0,1]$ and $Bin(n,\rho)$ is the binomial distribution. Alternatively, any
distribution with support $\{0,\ldots,d_{t-1}+1\}$ may be used.

\section{Algorithms}
\label{sec:algorithms}
The full posterior distribution described in the previous section is
intractable; we therefore present two algorithms, one which provides fully
Bayesian estimates of the regression coefficients, and the other which
provides approximate MAP estimates.

\subsection{Approximate MAP estimation}

MAP estimation requires maximization of the following objective function%
\begin{equation}
\sum_{t=1}^{T}\log p(y_{t}|\beta_{t})+\sum_{t=d}^{T}\sum_{j=1}^{p}\log
\pi(\beta_{j,t-d:t})-\sum_{t=d}^{T-1}\sum_{j=1}^{p}\log\pi(\beta
_{j,t-d+1:t})\label{eq:trueMAP}%
\end{equation}
This objective function is not convex and does not admit any latent variable
construction that might enable the use of an EM algorithm. We propose here an
online algorithm to perform approximate MAP\ estimation. The algorithm will
successively maximize $p(\beta_{t}|\widehat{\beta}_{1:t-1},y_{t})$ w.r.t.
$\beta_{t}$ for $t=1,\ldots,T$. At each time $t$, we therefore consider
optimization of the following objective function w.r.t. $\beta_{t}$%
\begin{equation}
\log p(\beta_{t}|\widehat{\beta}_{t-d:t-1})+\log p(y_{t}|\beta_{t}%
)\label{eq:approxMAP1}.
\end{equation}

It is easy to show that%
\begin{equation}
\beta_{j,t}|\beta_{j,t-d:t-1}\sim GH\left(  \alpha\beta_{j,t-1},\nu-\frac
{d}{2},\sqrt{\delta^{2}+\left\Vert \beta_{j,t-d:t-1}\right\Vert _{\Sigma_{d}%
}^{2}},\frac{\gamma}{\sqrt{1-\alpha^{2}}}\right)
\end{equation}
and we can therefore solve (\ref{eq:approxMAP1}) with an EM algorithm using
the scale mixture of Gaussian representation of the generalized hyperbolic distribution~\cite{Dempster1977,Caron2008}.

We now propose a second algorithm to obtain approximate MAP estimates.
Consider here that $\nu=(d+2)/2$, $d$ is fixed and known and $\delta=0$. We
can solve the following group lasso sliding window optimization problem at
time $t>d$, by maximizing according to $\beta_{t-d:t}$%
\begin{equation}
\log p(\beta_{t-d:t})+\log p(Y_{t-d:t}|\beta_{t-d:t})
\end{equation}
which reduces to minimizing
\begin{equation}
\frac{1}{2\sigma^{2}}\left\Vert Y_{t-d:t}-X_{t-d:t}\beta_{t-d:t}\right\Vert
^{2}+\gamma\sum_{j=1}^{p}\left\Vert \beta_{j,t-d:t}\right\Vert _{\Sigma_{d+1}},
\end{equation}
a convex group lasso problem~\citep{Yuan2006} for which efficient
algorithms exist.

\subsection{Sequential Monte Carlo algorithm}

While the previous algorithms conduct approximate MAP inference, we can also
write a sequential Monte Carlo (SMC) algorithm to conduct fully Bayesian
inference \citep*{Doucet2001}. Particularly, as memory requirements prevent implementing a particle filter with 1,000,000 particles, we employ the particle independent Metropolis-Hastings algorithm~\citep*{Andrieu2010} to approximate the
full posterior $\pi(\beta_{1:T},d_{1:T}|y_{1:T})$. We can use latent variables to produce efficient proposal
distributions for $\beta_{t}$
\begin{equation}
\text{for }j=1,\ldots,p,~\ \tau_{j,t}|\beta_{j,t-d_{t}:t-1},d_{t}\sim
GiGauss(\nu-d_{t}/2,\sqrt{1-\alpha^{2}}\sqrt{\delta^{2}+\left\Vert
\beta_{j,t-d_{t}:t-1}\right\Vert _{\Sigma_{d_{t}}}},\frac{\gamma}%
{\sqrt{1-\alpha^{2}}})\label{eq:latent}%
\end{equation}
and
\begin{equation}
\beta_{t}|\tau_{t},y_{t}\sim\mathcal{N}(\mu_{t},\Sigma_{t}%
)\label{eq:samplingbeta}%
\end{equation}
with $\mu_{t}=(\sigma_{t}^{2}D_{\tau_{t}}^{-1}+X_{t}^{\prime}X_{t}%
)^{-1}(\alpha\sigma_{t}^{2}D_{\tau_{t}}^{-1}\beta_{t-1}+X_{t}^{\prime}y_{t})$
and $\Sigma_{t}=(D_{\tau_{t}}^{-1}+X_{t}^{\prime}X_{t}/\sigma_{t}^{2})^{-1}$,
$D_{\tau}=diag(\tau)$. We sample from $\pi(\beta_{t}|\tau_{t},y_{t})$, and the
weights are simply updated with%
\begin{equation}
\pi(y_{t}|\tau_{t})=\mathcal{N}(y_{t};\alpha X_{t}\beta_{t-1},X_{t}^{T}D_{\tau_{t}%
}X_{t}+\sigma_{t}^{2}I_{n})\label{eq:weightsupdate}%
\end{equation}
where $\mathcal{N}(x;\mu,\Sigma)$ is the probability density function of the Gaussian distribution of mean $\mu$ and covariance matrix $\Sigma$ evaluated at $x$. The sequential Monte Carlo algorithm is described in Algorithm~\ref{algo:SMC}, and the particle independent Metropolis-Hastings algorithm in Algorithm~\ref{algo:PIMH}.

\begin{algorithm}[h!]
\caption{Sequential Monte Carlo algorithm}
\label{algo:SMC}
\underline{At $t=1$}

\qquad$\bullet$ For $i=1,\ldots,N$

\qquad\qquad$\bullet$ Set $d_{1}^{(i)}=0$

\qquad\qquad$\bullet$ For $j=1,\ldots,p$, sample $\tau_{j,1}^{(i)}\sim
GiGauss(\nu,\delta,\gamma)$

\qquad\qquad$\bullet$ Sample $\beta_{1}^{(i)}\sim\mathcal{N}((\sigma_{1}%
^{2}D_{\tau_{1}}^{-1}+X_{1}^{\prime}X_{1})^{-1}X_{1}^{\prime}y_{1}%
,(D_{\tau_{1}}^{-1}+X_{1}^{\prime}X_{1}/\sigma_{1}^{2})^{-1})$

\qquad\qquad$\bullet$ Compute the weights
\[
w_{1}^{(i)}=\pi(y_{1}|\tau_{1}^{(i)})
\]

\qquad$\bullet$ Replicate particles of high weights and delete particles of
low weights, so that to obtain a new set of particles.

\underline{For $t=2,\ldots$}

\qquad$\bullet$ For $i=1,\ldots,N$

\qquad\qquad$\bullet$ Sample $d_{t}^{(i)}\sim Bin(d_{t-1}^{(i)}+1,\rho)$

\qquad\qquad$\bullet$ For $j=1,\ldots,p$, sample $\tau_{j,t}^{(i)}\ $\ from
Eq. (\ref{eq:latent})

\qquad\qquad$\bullet$ Sample $\beta_{t}^{(i)}$ from Eq. (\ref{eq:samplingbeta})

\qquad\qquad$\bullet$ Compute the weights
\[
w_{t}^{(i)}=\pi(y_{t}|\tau_{t}^{(i)})
\]

\qquad$\bullet$ Replicate particles of high weights and delete particles of
low weights, so that to obtain a new set of particles.\bigskip
\end{algorithm}

\begin{algorithm}[h!]
\caption{Particle Independent Metropolis-Hastings algorithm}
\label{algo:PIMH}

\underline{Initialization}

\qquad$\bullet$ Run the sequential Monte Carlo algorithm~\ref{algo:SMC}, sample $$\widetilde\beta_{1:T}^{(1)}\sim \sum_{i=1}^N w_T^{(i)}\delta_{\beta_{1:T}^{(i)}}$$  and compute
$$
\widehat Z^{(1)}=\prod_{t=1}^T\left (\frac{1}{N}\sum_{i=1}^N  w_t^{(i)}\right )
$$

\underline{At iteration $m\geq 2$}

\qquad$\bullet$ Run the sequential Monte Carlo algorithm~\ref{algo:SMC}, sample $$\widetilde\beta_{1:T}^{*(m)}\sim \sum_{i=1}^N w_T^{(i)}\delta_{\beta_{1:T}^{(i)}}$$  and compute
$$
\widehat Z^{* (m)}=\prod_{t=1}^T \left (\frac{1}{N}\sum_{i=1}^N  w_t^{(i)}\right )
$$

\qquad$\bullet$ With probability
$$
1 \wedge \frac{\widehat Z^{* (m)}}{\widehat Z^{(m-1)}}
$$
set $\widetilde\beta_{1:T}^{(m)}=\widetilde\beta_{1:T}^{*(m)}$ and $\widehat Z^{(m)}=\widehat Z^{* (m)}$, otherwise set $\widetilde\beta_{1:T}^{(m)}=\widetilde\beta_{1:T}^{(m-1)}$ and $\widehat Z^{(m)}=\widehat Z^{(m-1)}$

\end{algorithm}

\section{Simulation Study}
\label{sec:simulation}
\begin{figure}[ptb]
\begin{center}
\includegraphics[width=.95\textwidth]{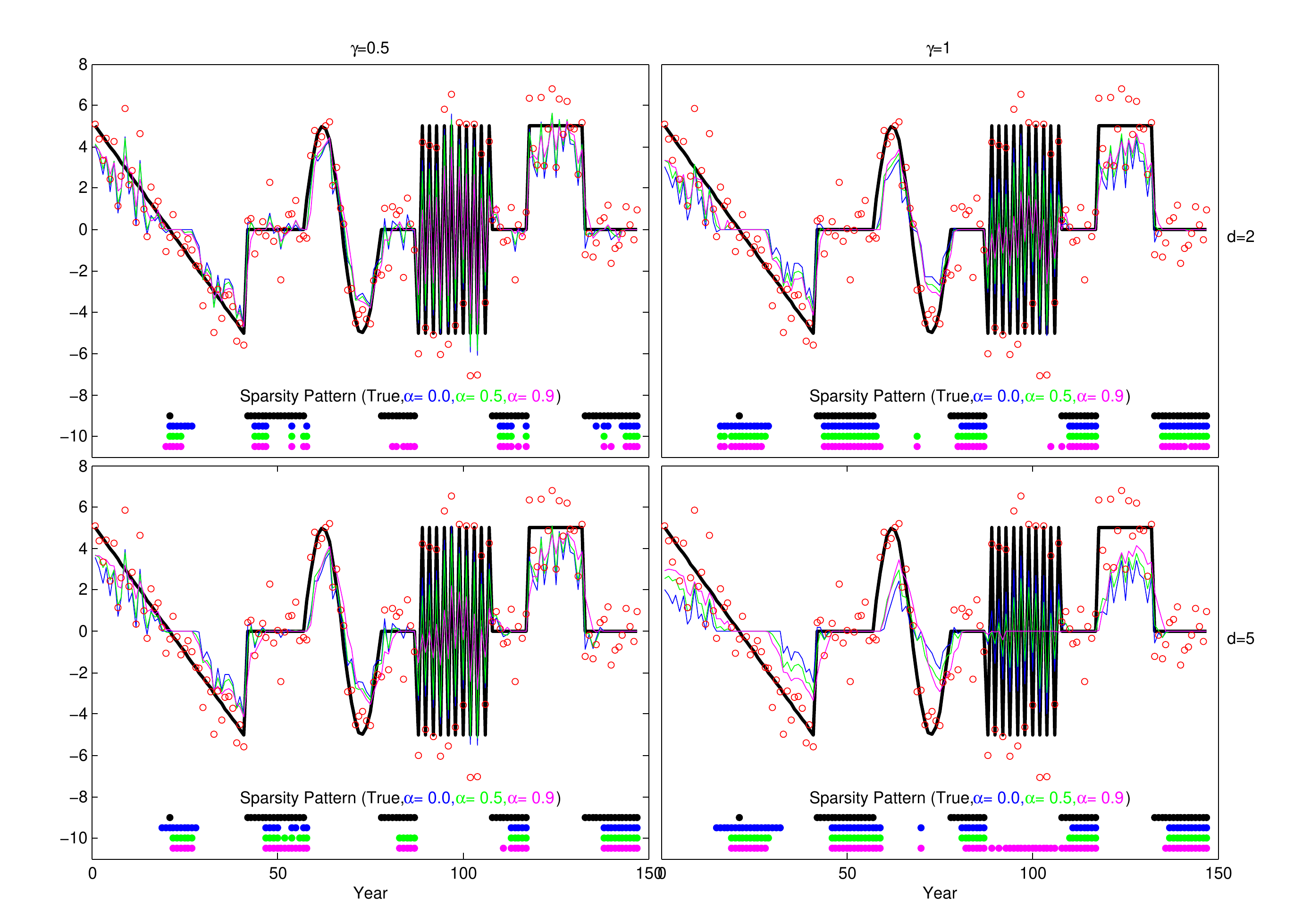}
\end{center}
\caption{Simulation Results. Ground truth (black line), observations (red
circles), and fitted models (blue line, $\rho=0$; green line, $\alpha=.5$;
magenta line, $\alpha=.9$). The bottom of each plot shows the estimated
sparsity pattern for the ground truth and estimates, with corresponding color
coding. We observe that shrinkage is controlled by $\lambda$ and smoothness in
the estimated sparsity pattern by $d$.}%
\label{fig:sim}%
\end{figure}

We now conduct a simulation study to explore the properties and performance of
the statistical model. We first generate an artificial time series of
``observations'' (red circles, Figure \ref{fig:sim}) generated from ground
truth (solid black line, Figure \ref{fig:sim}) with additive Gaussian noise.
We then explore the model's performance for $d=2,5$, $\nu= (d+2)/2$,
$\delta=0$, $\gamma=.5,1$, and $\alpha=0,.5,.9$. Figure \ref{fig:sim} shows
the model's MAP estimate of ground truth for each parameter setting, as well as
the true and estimated sparsity patterns. We immediately notice that the model provides shrinkage in the
estimates, as controlled by $\gamma$. In addition, the ability of the model to
detect sparsity is dependent on the correlation structure. For example, when
the ground truth consists of alternating values $+5$ and $-5$, and $\alpha=
.9$, the model fits this section as being sparse, due to the lack of
smoothness. We also notice that the parameter $d$ has two major implications.
Firstly, it creates smoothness and stability in the estimate of sparsity
structure. Secondly, because the model is fit online and hence the model
is in some sense a filter, there is a slight delay in detecting sparsity
patterns, the size of which increases with $d$.

\begin{figure}[ptb]
\begin{center}
\includegraphics[width=.5\textwidth]{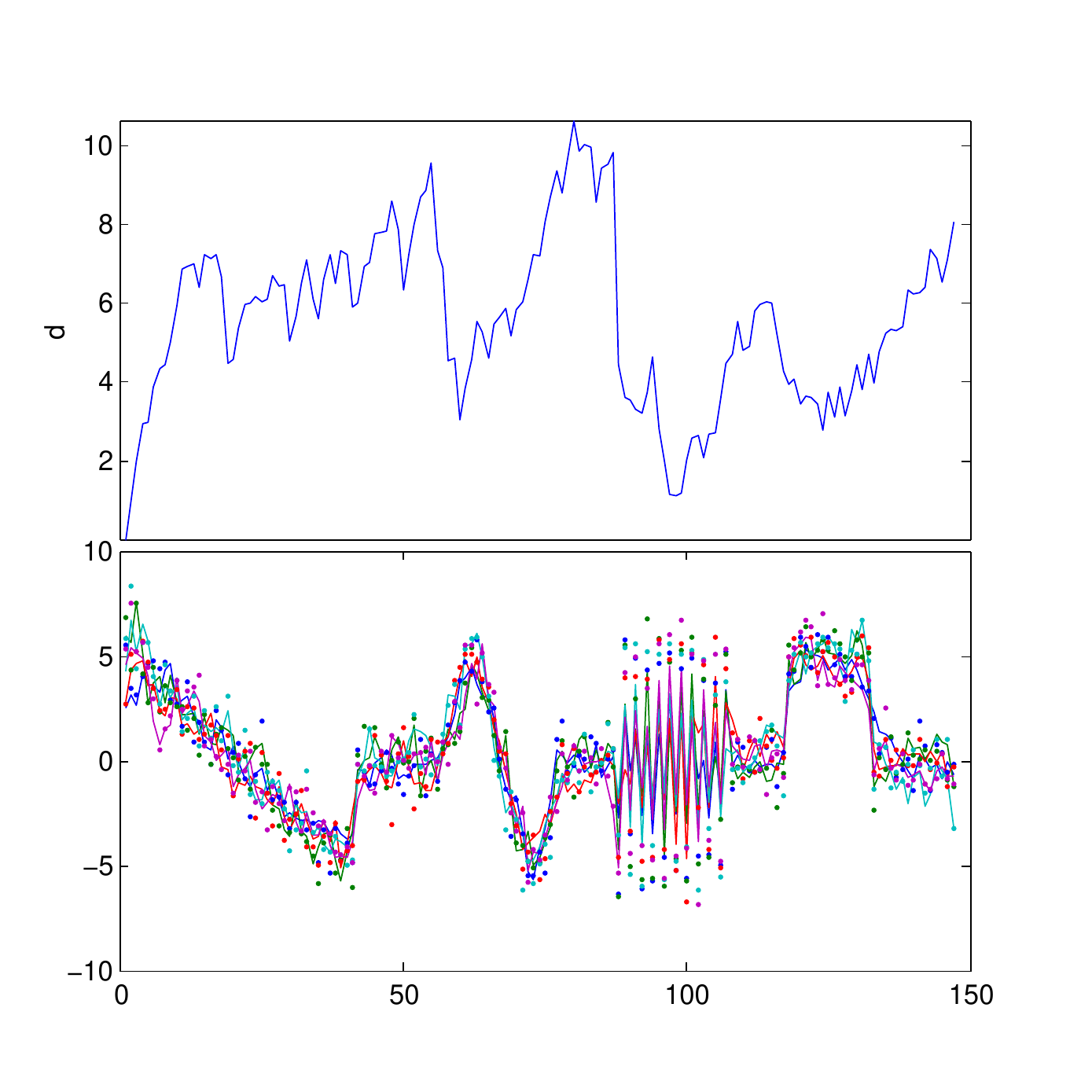}
\end{center}
\caption{Results from fitting simulation data using SMC. Estimate of $d$ over
time (top). Simulated observations and fitted model for $5$ replicates
(bottom).}%
\label{fig:sim_pmcmc}%
\end{figure}

We now sample from the posterior distribution through the
aforementioned sequential Monte Carlo algorithm, using $1000$ iterations of
the particle independent Metropolis-Hastings algorithm \citep*{Andrieu2010},
each with $1000$ particles. We set $\sigma_{t}$, $\nu$, and $\gamma$ to $1$,
and $\delta$ to $0.01$. Also, to induce moderate correlation, we select
$\alpha= 0.8$, and for the temporal correlation in $d$, set $\rho= 0.9$. These
choices were made to demonstrate the estimation of $d$, although we emphasize
that depending on the circumstances and model criterion other parameter
choices provide wide modeling flexibility, allowing practitioners to recreate
several models in the literature \citep*{Snoussi2006,Griffin2007,Caron2008,
Griffin2010}, as well as build unique models which extend and
bridge between these models. Figure \ref{fig:sim_pmcmc} shows the resulting
inference for $5$ replicates of the model. Here we see that the estimate of
$d$ ranges between $2$ and $10$, dropping during time steps when the structure
of the simulated time series changes. We also plot the fitted model for each
replicate, where we observe that while the fully Bayesian model does not
produce sparsity, it does induce shrinkage and smoothing of the process.

\section{Modelling Stock Volatility}
\label{sec:stockvolatility}

Stocks, as well as their derivatives, are known to alternative periods of
stability and change, both on a micro and a macro scale. On a micro scale,
this often occurs due to a news item or press release setting off a flurry of
trading of a given asset. As an example, consider the stock price of BP oil
and gas company following news of the Deepwater Horizon oil spill in 2010.
Following this news, the regular day-to-day variability in the stock price
increased by orders of magnitude as a constant stream of good and bad news led
to an increase in trading activity. On a macro scale, this is often due to
crashes, or corrections, in the market. As an example, consider the 2000 tech
bubble, or the 2008 stock market crash, which both led to massive changes in
the stock market as a whole.

Stock price time series are freely available from numerous sources such as
Yahoo! and others, and we study a collection of stock and derivatives which we
suspect would exhibit interesting effects in their prices over the period 1998
to 2011. The first stock we study is BP; as mentioned earlier, we expect to
observe massive variability following the 2010 oil spill. We similarly follow
OIL, iPath's S\&P Crude Oil Index. Conversely, we look at PowerShares' Crude
Oil Short (SZO), to study the effect of shorting the price of crude oil. The
next asset we study is XCI, the Amex Computer Technology Index, in the hopes
of observing activity from the 2000 tech bubble. Next we turn to the real
estate market, as measured through ProLogis (PLD), a real estate investment
trust which began in 2006, with particular interest in the 2008 market crash.
These five stocks, indexes, and derivatives constitute the core of our study.

\begin{figure}[ptb]
\begin{center}
\includegraphics[width=.6\textwidth]{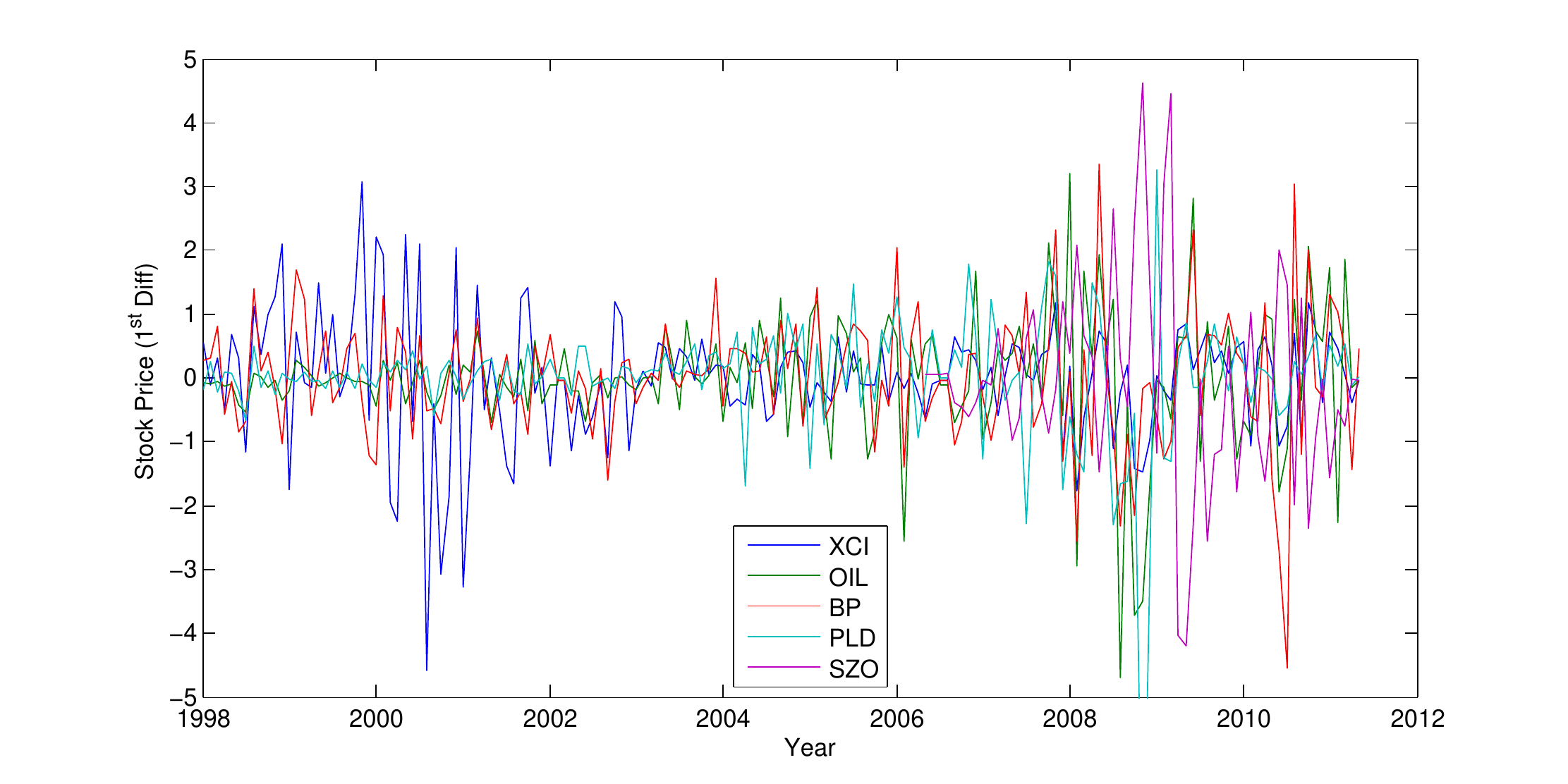}
\end{center}
\caption{First differences of a portfolio of stocks and derivatives. Note the
similarities with realizations from the model (Figure \ref{fig:samples}). From
this plot, it is difficult to decipher major trends from noise. In contrast,
see Figure \ref{fig:stockmodel}.}%
\label{fig:stockdata}%
\end{figure}

We begin by plotting the monthly change in the previously mentioned stocks and
derivatives in Figure \ref{fig:stockdata}, where we see significant volatility
in the technology sector in the early 2000's, and similar volatility in all
sectors during the 2008 recession and recovery. One pattern of immediate note is that of SZO, the PowerShares'
Crude Oil Short, which as expected reacts contrary to the other assets during
the 2008 recession.

\begin{figure}[ptb]
\begin{center}
\includegraphics[width=.95\textwidth]{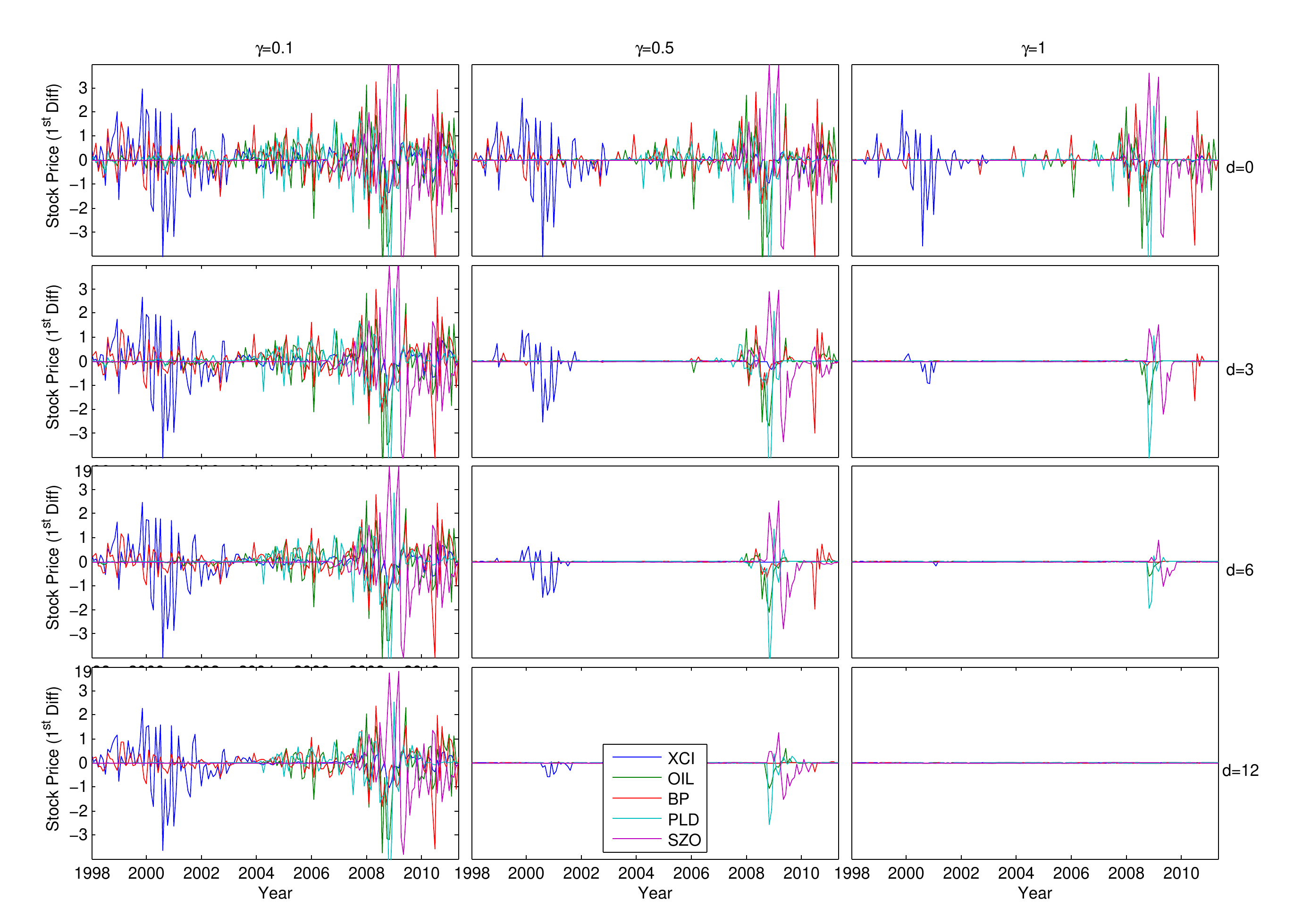}
\end{center}
\caption{Model fit of first differences of a portfolio of stocks and
derivatives. Note the effects of the early 2000's tech bubble in the
technology index (XCI), the 2008 housing market crash in the real estate
investment trust (PLD), and the 2010 oil spill on the BP stock.}%
\label{fig:stockmodel}%
\end{figure}

We now attempt to model these volatilities directly, using parameters $d=0,3,6,12$, $\nu= (d+2)/2$,
$\delta=0$, $\gamma=0.1,.5,1$, and $\alpha=.5$. Figure
\ref{fig:stockmodel} plots the filtered time series for the chosen ranges of $\gamma$ and
$d$, where we see that for moderate values of these parameters (namely, the
two center panels), we are able to isolate significant events, such as the
early 2000's tech bubble and the 2008 recession, particularly in the housing
market. 

\begin{figure}[ptb]
\begin{center}
\includegraphics[width=.95\textwidth]{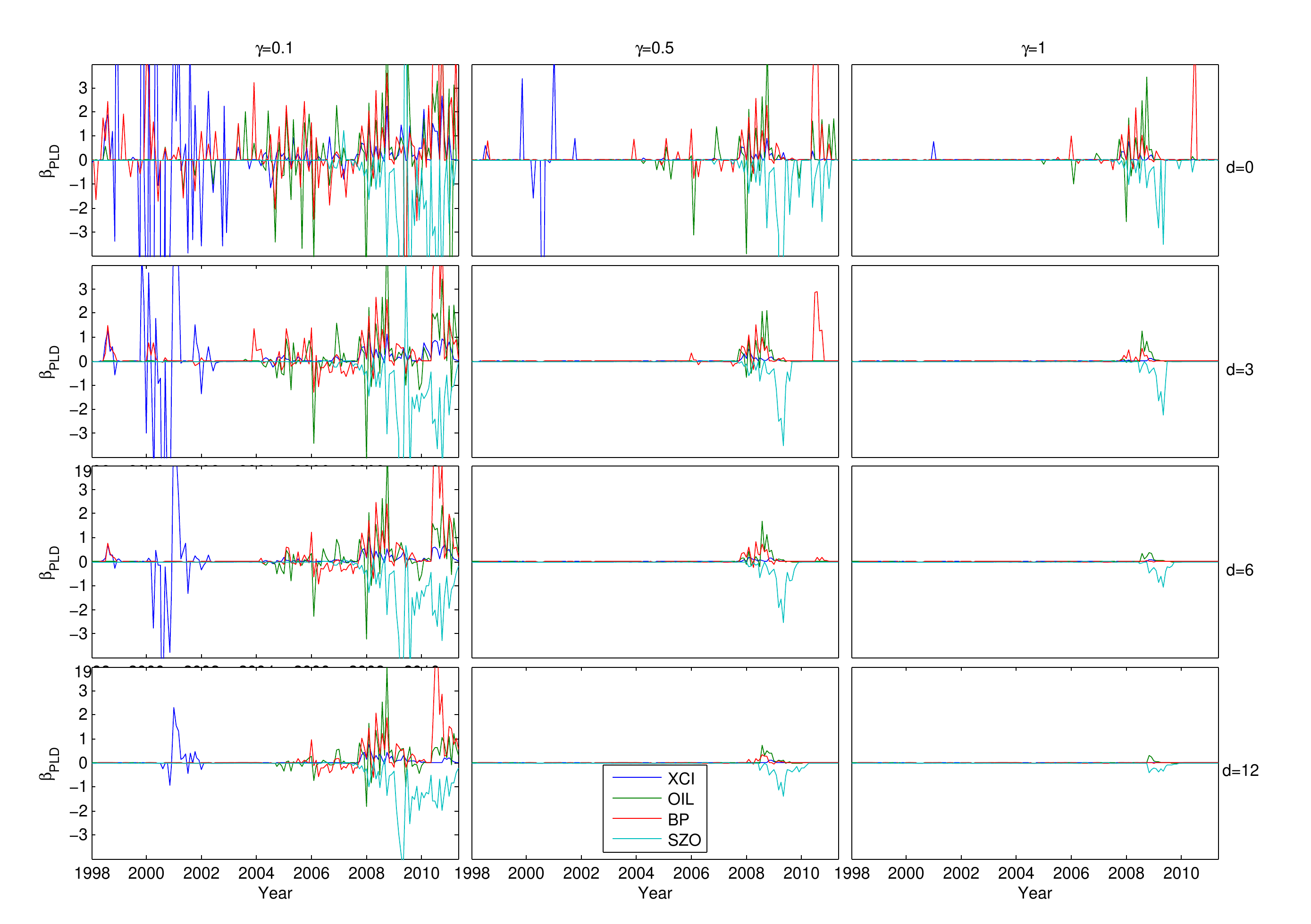}
\end{center}
\caption{Regression coefficients of a portfolio of stocks and derivatives
regressed on PLD. Note that as $d$ increases, the estimation of the
coefficients is stabilized.}%
\label{fig:stockmodel2}%
\end{figure}

Taking a closer look at the real estate market, we now consider the very
practical problem of building a portfolio in the situation where one is
already largely invested in the real estate market, namely through home
ownership. Specifically, the casual investor who owns their own home and
wishes to diversify should aim to build a portfolio with little correlation to
the housing market in case of another housing crisis. Modifying the problem
slightly to regress the remaining four assets against PLD, the housing index,
we calculate the regression coefficients $\beta_{PLD}$ for each time series,
plotting them in Figure \ref{fig:stockmodel2}. We note immediately that for small $\gamma$ and $d$ (the top left
panel), it appears that all assets are highly correlated with the housing
market. As we induce shrinkage and sparsity, however, we notice that the
technology index, XCI, disappears. As expected, the derivative which is short
on oil (SZO) is negatively correlated with the housing market during the
housing collapse in 2008. Conversely, the remaining two variables are
positively correlated during the crash. Using these results, one might choose
to diversify their investment in their house with technology stocks, or as an
alternative to hedging the housing market directly might instead hedge the
price of crude oil.

\section{Discussion and Extensions}
\label{sec:discussion}

The dependency structure of the $\beta_{j,1:T}$ is a d-order Markov model, which is a decomposable graph structure. The construction proposed in this paper could be generalized to any dependence structure that is given by a decomposable graph~\citep*{Lauritzen1996}, where the joint distribution on cliques of the graph is generalized hyperbolic. This would enable one to consider dependencies, for example, on rooted trees.\bigskip

In this paper, we have proposed a novel approach for conducting dynamically sparse Bayesian regression.  Built on the class of multivariate generalized hyperbolic distributions, the proposed method generalized many existing approaches for tackling this problem, while providing added modeling flexibility.  Inference on this class of models may be conducted exactly using MCMC methods, in particular the particle independent Metropolis-Hastings algorithm, or approximate but sparse approximations may be built around MAP estimator, using overlapping group lasso techniques or the EM algorithm.

The proposed class of models is well-suited to modeling stock volatility data, as the structure of the multivariate generalized hyperbolic distribution induces alternating periods of large and small volatility as observed daily market fluctuations.  We demonstrate how, through this class of models, one is able to isolate large-scale variation in stock price volatility to build a conservative and robust portfolio of uncorrelated assets.

\singlespacing
\bibliographystyle{apalike}
\bibliography{tvvarsel}

\end{document}